\documentstyle[prb,aps,floats,psfig]{revtex}
\psfigurepath{/homes/solid/radtke/modes}

\begin{document}
\twocolumn[
\hsize\textwidth\columnwidth\hsize\csname@twocolumnfalse\endcsname

\title{Collective Modes in a Symmetry-Broken Phase:\\
Antiferromagnetically Correlated Quantum Wells}
\author{R. J. Radtke and S. Das Sarma}
\address{Center for Superconductivity Research, Department of Physics,\\
University of Maryland, College Park, Maryland  20742-4111}
\maketitle

\begin{abstract}
We investigate the intersubband spin-density-excitation spectrum
of a double quantum well in a low-density symmetry-broken phase
with interwell antiferromagnetic correlations.
This spectrum is related to the intensity measured in depolarized inelastic
light scattering (ILS) experiments and therefore provides a means
of empirically identifying the antiferromagnetic phase.
Our computations reveal the existence of two collective modes,
a damped Nambu-Goldstone (NG) mode arising from the broken spin
symmetry and an undamped optical mode.
Since the NG mode contains most of the spectral weight,
ILS experiments will need to examine the low-frequency response
for signatures of the antiferromagnetic phase.
\\
\end{abstract}



]

Electronic correlations become increasingly important as
either the dimensionality or the effective electron density is lowered.
These enhanced correlations may give rise to new states of matter,
of which the best known example is the fractional quantum Hall
liquid.\cite{fqhe}
This phase occurs in 2D electron gases in high magnetic fields,
but correlation-induced quantum phase transitions should also occur in
sufficiently pure low-dimensional, low-density quantum systems in
the {\it absence} of magnetic fields.
At very low densities, energetic considerations suggest that
the electron liquid in these systems will crystalize to form a Wigner
solid,\cite{Wigner} and this transition has been observed in a
classical 2D electron gas on the surface of liquid helium.\cite{Grimes}

The possibility of other novel quantum phases in the density regime
between the Wigner crystal and the electron liquid has been explored
both experimentally and theoretically in several recent
publications.\cite{MacDonald,Varma,Ruden,Neilson,Katayama,Pablo,RS}
These publications consider effectively multilayer systems such
as double quantum wells (DQWs) or wide single quantum wells, in which
the layer or subband index acts as an additional degree of freedom.
Of particular current interest are correlation-driven transitions
resulting in a spontaneous charge transfer between the layers, which
is expected theoretically whether a magnetic field is
present\cite{MacDonald,Varma} or not\cite{Ruden}
and which may have been observed experimentally.\cite{Katayama}

Another correlation-induced phase transition, which awaits experimental
verification, involves the development of nontrivial, subband-coupled
antiferromagnetic order between the layers in certain two-layer quantum
well geometries in zero field.\cite{Pablo,RS}
The entrance into this antiferromagnetic phase is marked by the
collapse of the usual intersubband spin-density excitations (SDEs) as
the density is lowered.\cite{Pablo}
These excitations can be measured by depolarized inelastic light
scattering (ILS),\cite{Decca} and so the approach to the
antiferromagnetic phase boundary as a function of density should be
detectable.\cite{Plaut}
Nonetheless, unambiguous identification of this phase
would require the observation of the modified elementary
excitations\cite{Plaut} of the symmetry-broken phase, as occurred
in the Wigner solid with the detection of its phonons.\cite{Grimes}

In this paper, we assume uncritically the existence of the antiferromagnetic
phase and calculate its elementary excitations in the intersubband
spin-density channel as a guide for future experiments.
Since the antiferromagnetic ground state is not reproducible within
the dynamical spin-polarized local-density-approximation approach
underlying the original calculations,\cite{Pablo} we start from a
simple mean-field theory\cite{RS} and compute the collective
modes in a conserving approximation.\cite{conserving}
These calculations yield an intersubband response which is dominated
at low frequencies by the Nambu-Goldstone (NG) mode\cite{NG}
associated with the broken spin-rotation invariance of the antiferromagnetic
phase.
Thus, ILS experiments will need to examine the low-frequency portion
of the response for the signatures of this phase.
We believe that the results we present transcend the specific details of
our model and are, to the best of our knowledge, the first determination
of the collective modes of a low-density, double-quantum-well-type
structure in a symmetry-broken phase.

We consider an interacting electron gas subject to a confining potential
along one direction which results in a bimodal distribution of the
self-consistent charge density.
The antiferromagnetic phase may arise when the two layers are sufficiently
close that the lowest two subbands are well separated in energy from
the higher subbands.\cite{Pablo,RS}
At the temperatures and densities relevant for this study, we may
therefore neglect the higher subbands.
Furthermore, since the antiferromagnetic phase is driven by the
Coulomb interaction $V ({\bf r})$, we take a simple parameterization
of this interaction in order to obtain the generic features of this phase:
$V({\bf r}) = V \delta({\bf r})$.
Even at this very simple level of approximation, ferromagnetic phases
may arise at low densities due to the intrasubband matrix elements of
$V({\bf r})$.
Since the aim of this paper is to obtain the collective modes of the
antiferromagnetic phase and not to establish its existence,
we may set the intrasubband matrix elements to zero without loss
of generality.

\begin{figure}[tb]
\psfig{figure=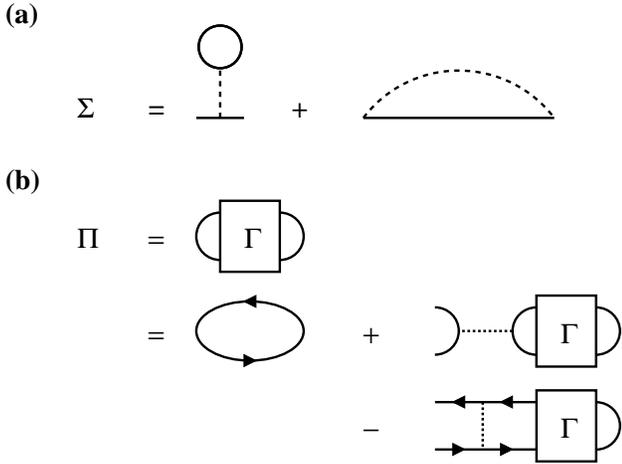,width=0.95\linewidth}
\caption{Many-body diagrams for (a) the self-energy $\Sigma$ and
(b) the polarization $\Pi$ used to compute the collective modes
discussed in the text.
The solid lines correspond to dressed electronic Green's functions
and the dashed lines to the effective interaction.
In evaluating these diagrams, we take the interaction to be
an on-site, repulsive interaction which makes
the integral equation in (b) tractable.}
\label{fig:diagrams}
\end{figure}

\begin{figure}[tb]
\psfig{figure=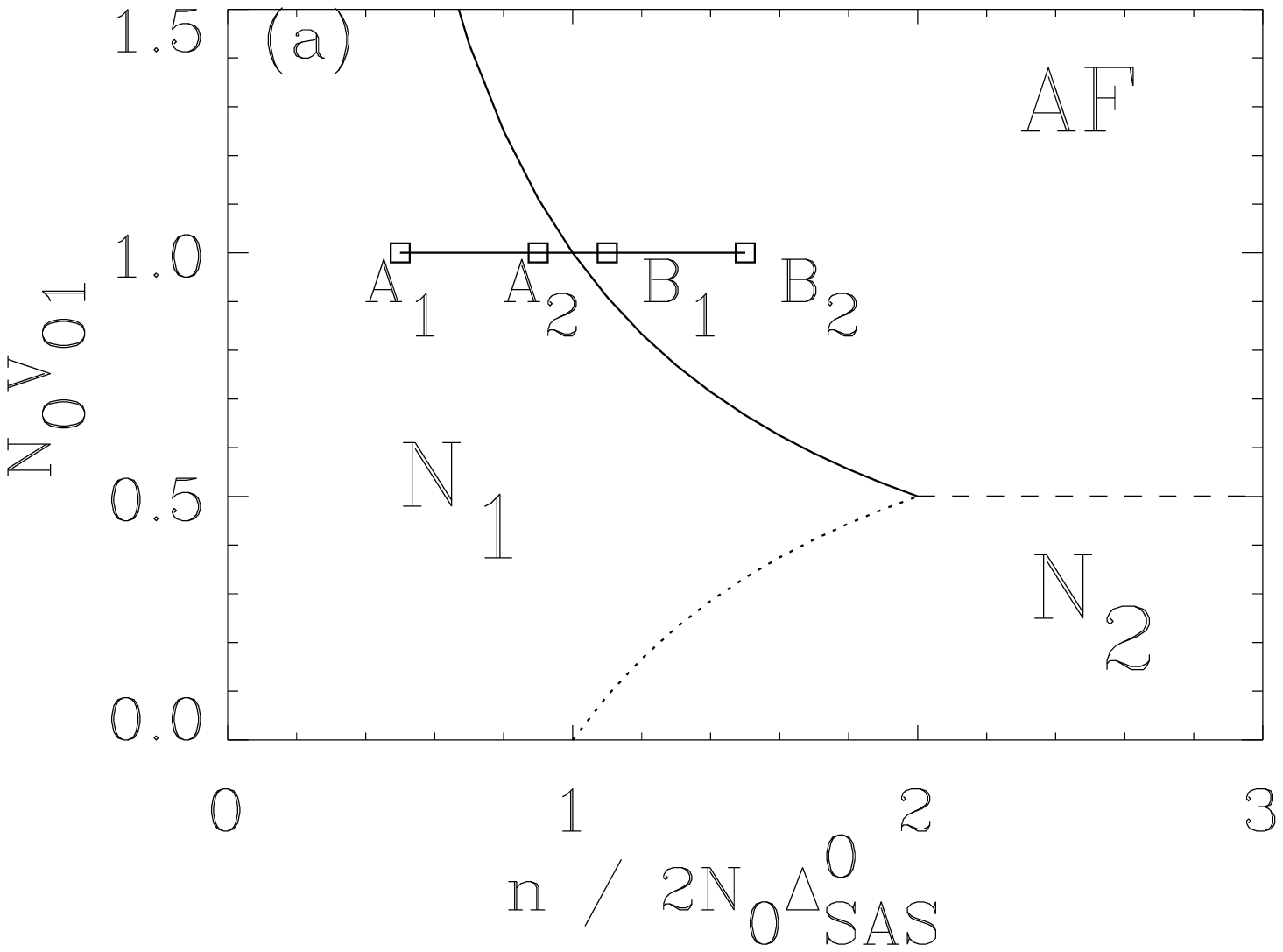,width=0.95\linewidth}
\psfig{figure=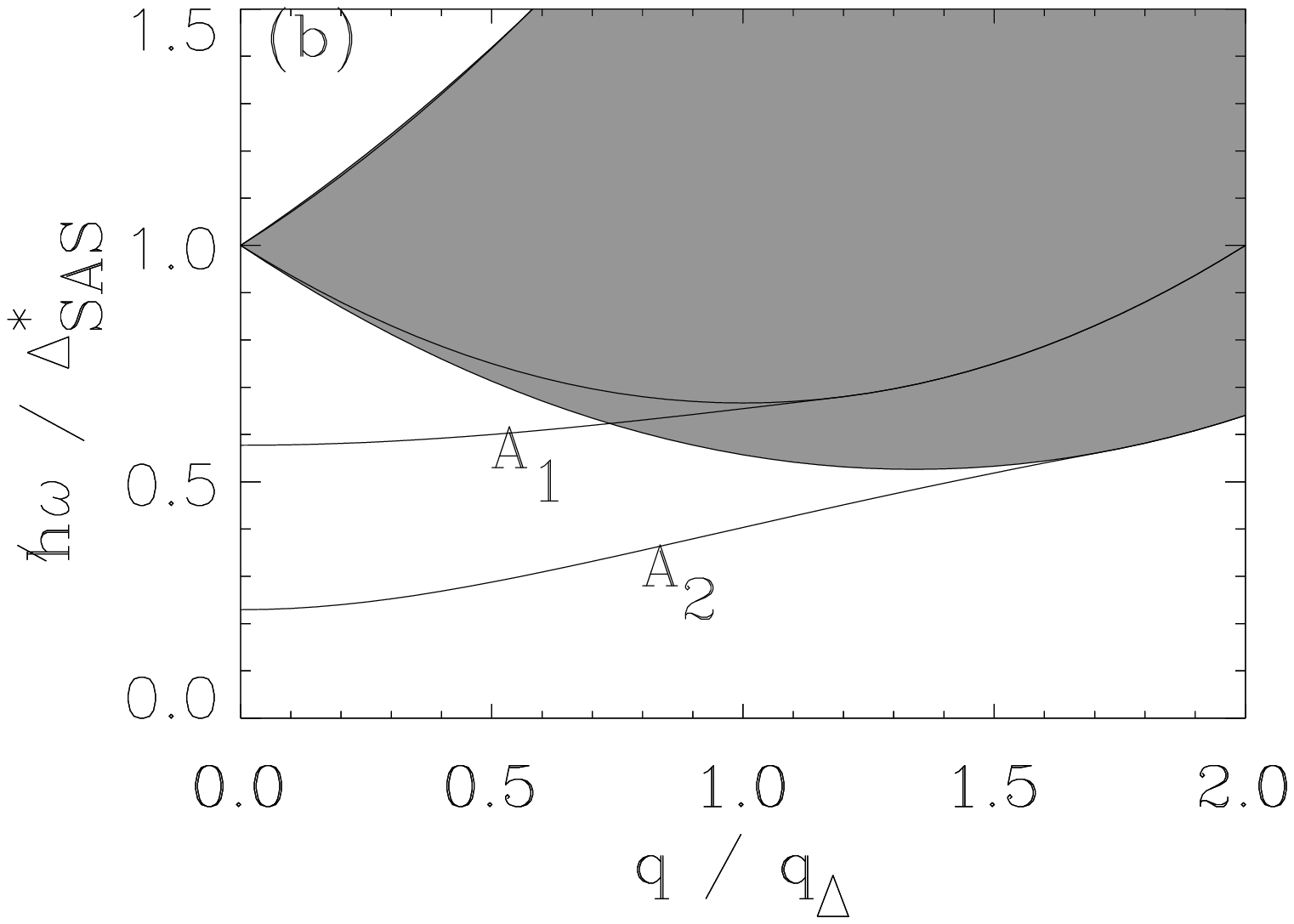,width=0.95\linewidth}
\caption{(a) Zero-temperature mean-field phase diagram for the
quantum well system described in the text as a function of
the intersubband Coulomb matrix element $V_{01}$
and the electronic density $n$.
The regions correspond to the normal (paramagnetic) phase with one
subband occupied ($N_1$), the normal phase with both subbands occupied
($N_2$), and the antiferromagnetic phase (AF).
The points labeled $A_1$, $A_2$, $B_1$, and $B_2$ are for future
reference.
Adapted from Ref.~\protect\onlinecite{RS}.
(b) Frequency $\omega$ of the intersubband spin-density excitations
as a function of wave vector $q$ (solid lines) in the normal phase
at the points $A_1$ and $A_2$ in (a).
The shaded area represents particle-hole excitations which damp
the collective modes.
In these figures, $N_0 = m / 2\pi \hbar^2$ is the single-spin,
2D density of states, $\Delta_{\rm SAS}^0$
is the splitting of the lowest two subbands when $V_{01} = 0$,
$\Delta_{\rm SAS}^*$ is the renormalized subband splitting, and
$q_{\Delta}^2 = 2m \Delta_{\rm SAS}^{0} / \hbar^2$.
}
\label{fig:phase}
\end{figure}

The self-energy for this simplified model is computed within
self-consistent Hartree-Fock theory, which is illustrated by
the diagrams in Fig.~\ref{fig:diagrams}(a).
In the subband representation, the antiferromagnetic phase
corresponds to a mixing of the non-interacting states
with opposite spin and different subband index.\cite{RS}
Thus, we look for solutions to the self-energy equations with
non-zero off-diagonal elements.\cite{note1}
Such solutions exist and are stable in the range of parameters
denoted by ``AF'' in Fig.~\ref{fig:phase}(a).
For a more detailed discussion of this phase diagram and the
self-energy and model underlying it, see Ref.~\onlinecite{RS}.

We compute the collective modes in the usual way\cite{SDS}
within a conserving approximation for the electronic
polarizability [cf. Fig.~\ref{fig:diagrams}(b)].\cite{conserving}
One complication that arises in this calculation is that the
mixing of the wave functions from different subbands in the antiferromagnetic
phase causes the ``bare'' polarizability to become off-diagonal.
Hence, the resulting polarizability matrices have
$16 \times 16$ elements, making analytic work nearly impossible.
However, the specific form of the interaction we use
allows a straightforward numerical solution for the interacting
polarizability through a matrix inversion of the equations.
The collective modes can then be obtained from the study of the
resulting interacting polarizability.
Since the intersubband spin-density excitations were the first sign
of the antiferromagnetic phase,\cite{Pablo} we restrict attention to
these excitations\cite{Plaut} in what follows and perform all our
calculations at zero temperature.

Before proceeding to our results, a few remarks on our
calculational scheme are in order.
At the level of the self-energy, we know that the Hartree-Fock
approximation is a poor one for the interacting electron gas
because it neglects screening effects.
Including these effects realistically is a difficult problem which
has not yet been solved.
In computing the collective modes, this problem is amplified by
the distinction one should draw between the interaction between
the bubbles, which is unscreened, and the interaction within the
bubbles, which should be screened.
Summing a particular set of screening diagrams may reduce the
error introduced into the self-energy, but would render the
collective mode calculation completely intractable.
Thus, we adopt a strong approximation to the actual Coulomb
interaction, a point-contact interaction, which should reproduce
the approximate qualitative features of the screening effects while
leaving a solvable set of equations.

To get a clear picture of the effect on the collective modes
of the transition from the normal paramagnetic phase
to the antiferromagnetic phase, we present calculations for four
representative points in the parameter space of our model.
These points are labeled in Fig.~\ref{fig:phase}(a) and
sample both the normal phase with one subband occupied
and the antiferromagnetic phase.
The two phases are separated by a second-order phase transition
at zero temperature.

In the normal phase [points $A_1$ and $A_2$ in Fig.~\ref{fig:phase}(a)],
a single, collective spin-density excitation exists below the energy of the
renormalized subband splitting, as shown by the solid lines in
Fig.~\ref{fig:phase}(b).
This mode disperses into the continuum of intersubband particle-hole
excitations denoted by the shaded areas in Fig.~\ref{fig:phase}(b), where
it becomes damped.
As one approaches the phase transition from the normal side
($A_1 \rightarrow A_2$), the
$q = 0$ frequency of the spin-density excitation $\omega_0$
decreases according to the relation
$\omega_0 / \Delta_{\rm SAS}^* = \sqrt{ 1 - V_{01} n / \Delta_{\rm SAS}^*}$,
where $\Delta_{\rm SAS}^* = \Delta_{\rm SAS}^0 + V_{01} n / 2$
is the renormalized subband splitting, $n$ is the electronic density,
$V_{01}$ the intersubband Coulomb repulsion,
and $\Delta_{\rm SAS}^0$ is the $V_{01} = 0$ subband splitting.
We see that the $q = 0$ spin density excitation softens completely
($\omega_0 \rightarrow 0$) when $V_{01} n \rightarrow 2\Delta_{\rm SAS}^0$.
This signals the onset of the antiferromagnetic phase.

\begin{figure}[tb]
\psfig{figure=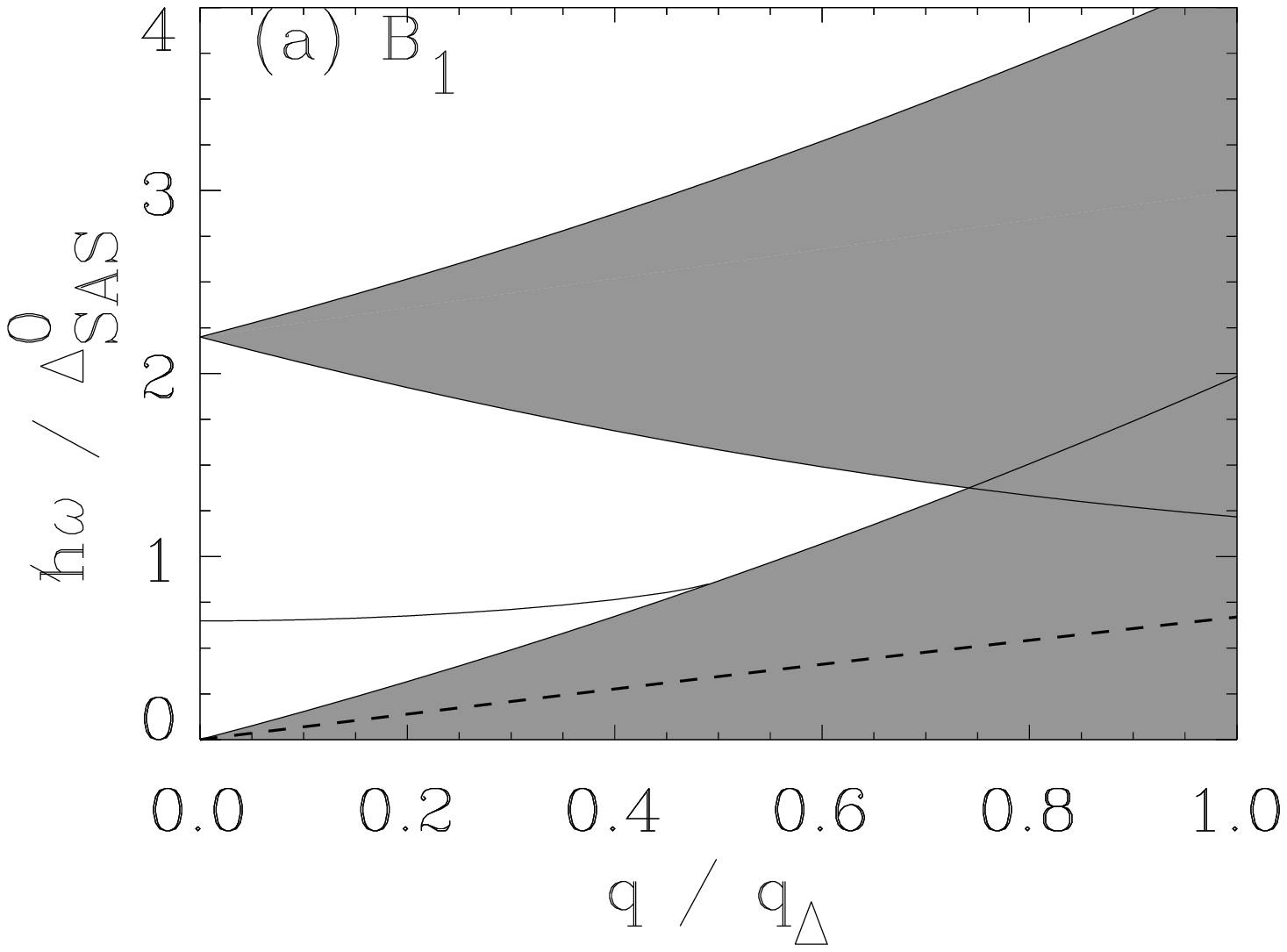,width=0.95\linewidth}
\psfig{figure=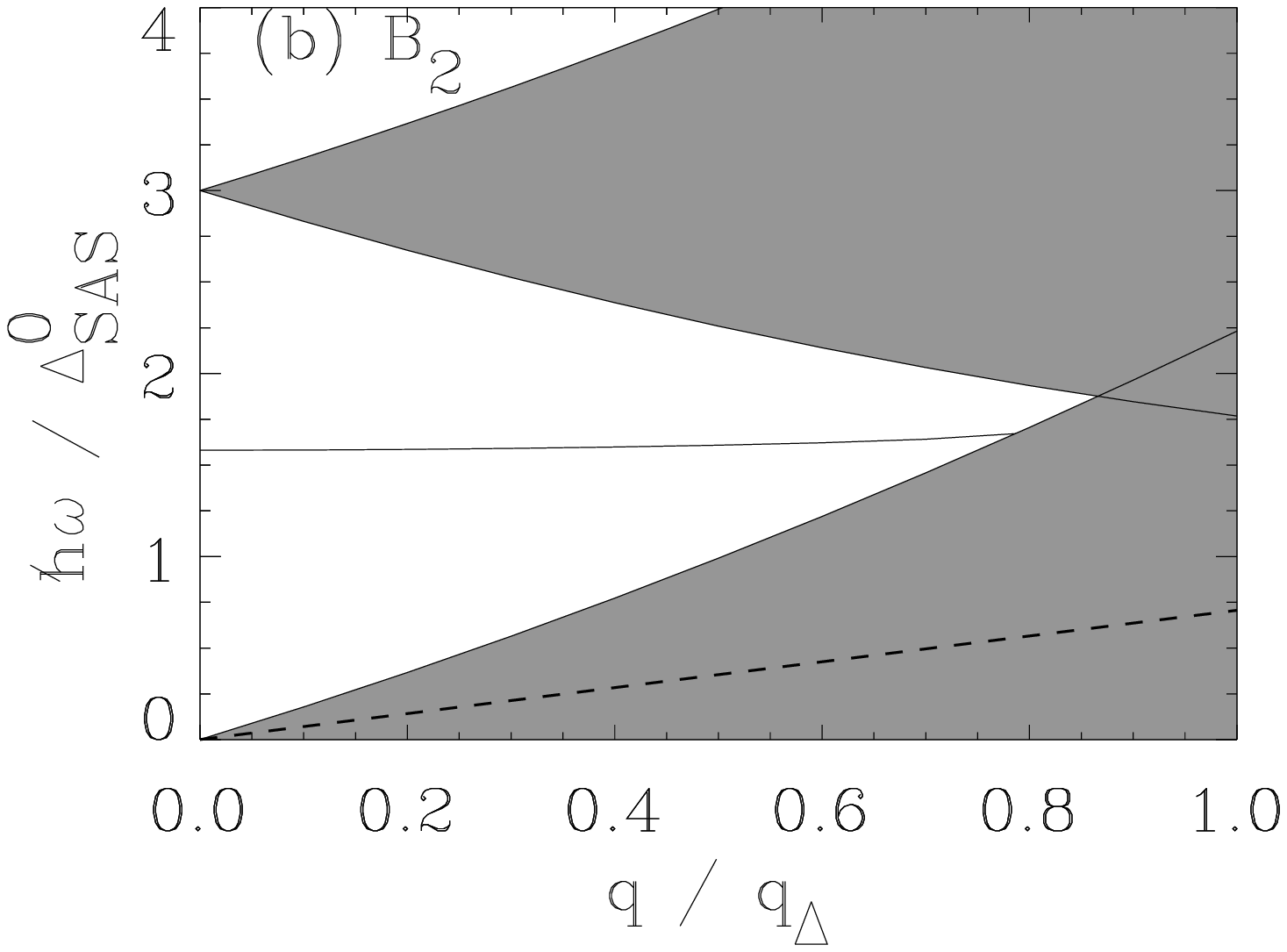,width=0.95\linewidth}
\caption{Frequency $\omega$ of the intersubband spin density excitations
as a function of wave vector $q$ in the antiferromagnetic phase
of the quantum well system discussed in the text
at the points (a) $B_1$ and (b) $B_2$ from Fig.~\protect\ref{fig:phase}(a).
The shaded area represents particle-hole excitations which damp
the collective modes.
Note that one optical mode (solid line) and
one damped acoustic mode (dashed line) are generically present and that
the intrasubband particle-hole continuum enters into the spectrum
due to the mixing of the single-particle wave functions in
the antiferromagnetic phase.
The notation is the same as in Fig.~\protect\ref{fig:phase}.}
\label{fig:qw}
\end{figure}

\begin{figure}
\psfig{figure=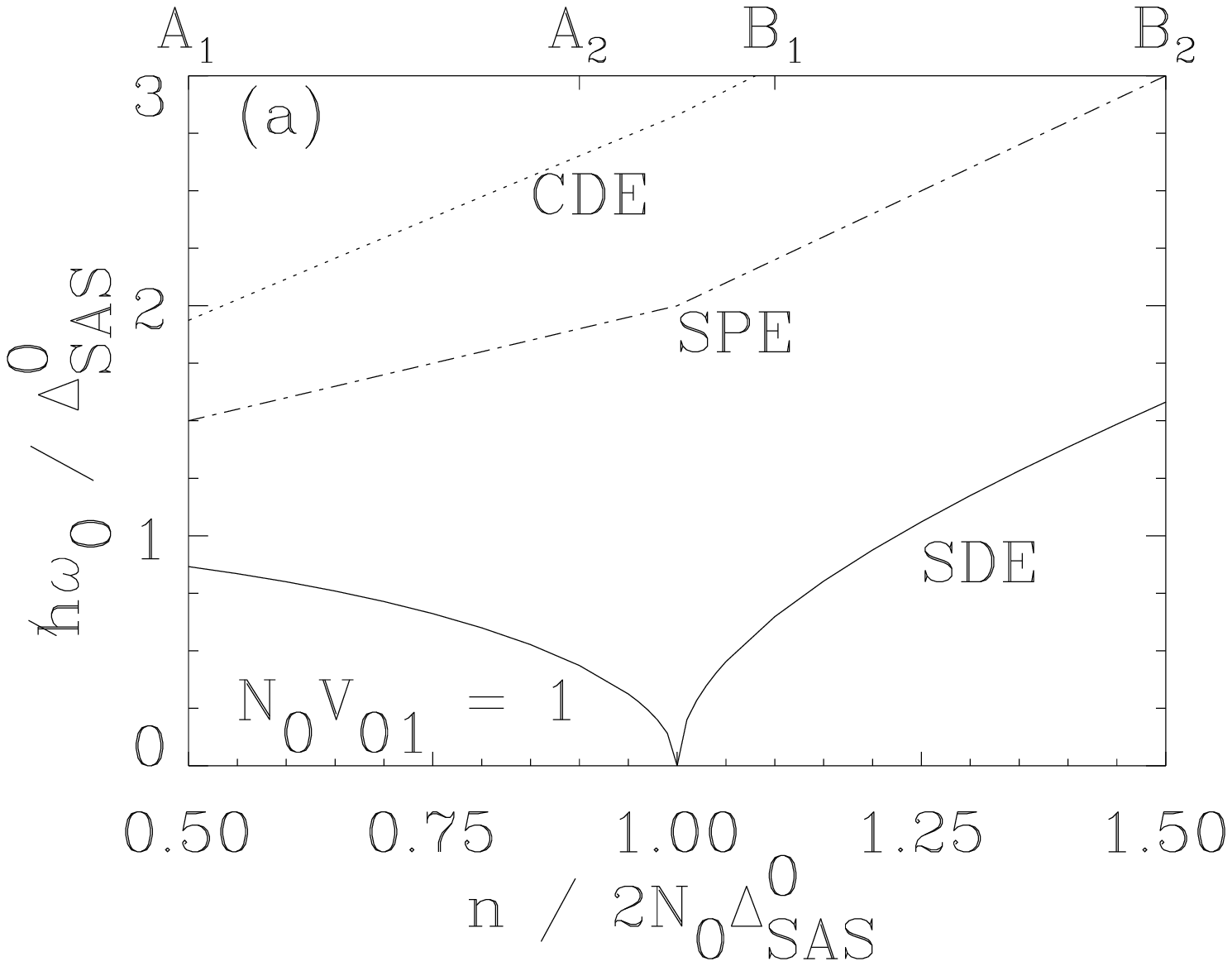,width=0.95\linewidth}
\psfig{figure=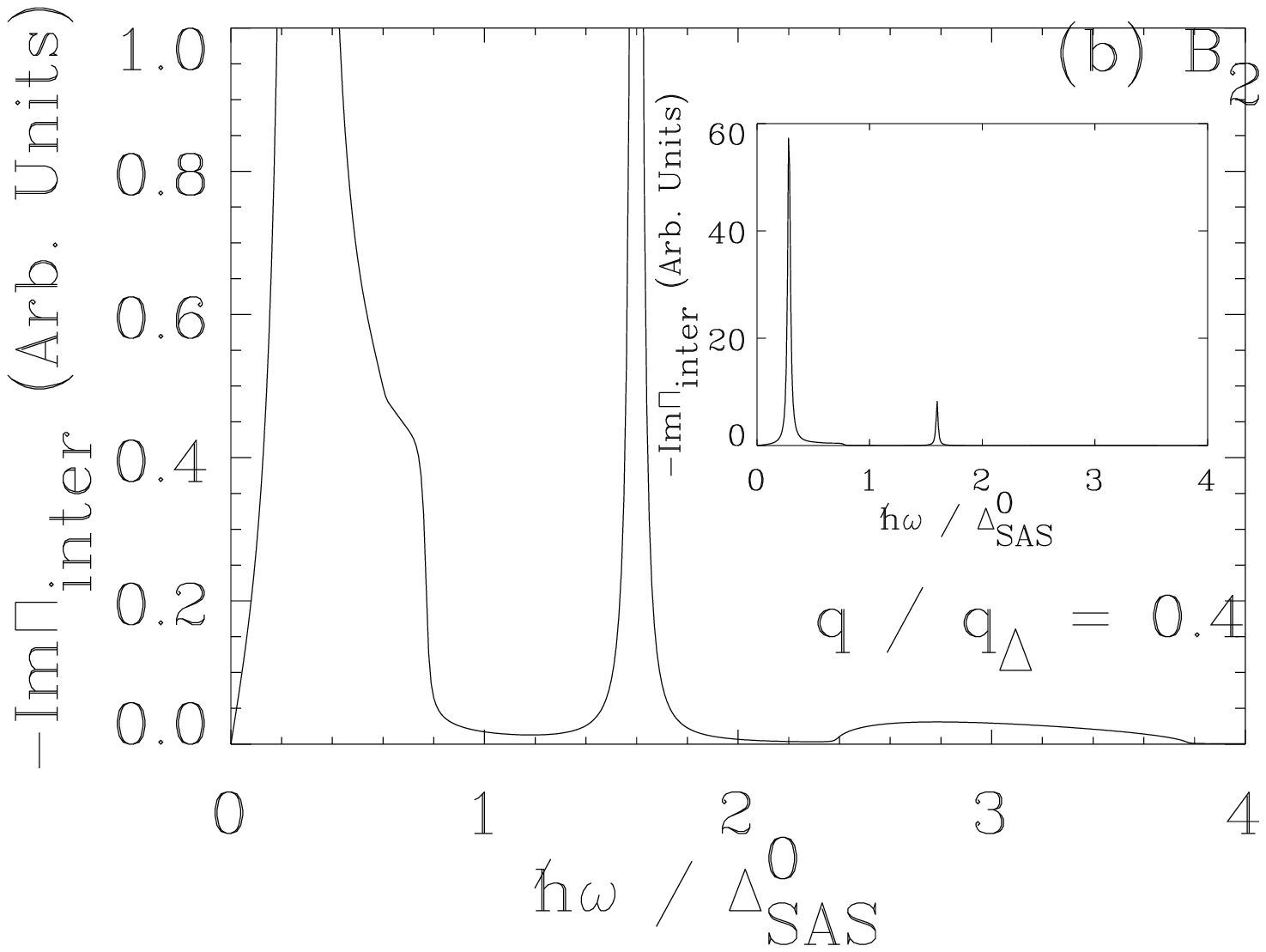,width=0.95\linewidth}
\caption{(a) Frequency at $q = 0$, $\omega_0$, of the intersubband
spin- (SDE, solid line) and charge-density excitations (CDE, dotted line)
as a function of electronic density $n$ for $N_0 V_{01} = 1$.
Also shown is the renormalized subband splitting
$\Delta_{\rm SAS}^*$ (SPE, dot-dashed line).
The top axis marks the points from the phase diagram in
Fig.~\protect\ref{fig:phase}(a).
Note the re-emergence of the SDEs in the antiferromagnetic
phase ($n / 2 N_0 \Delta_{\rm SAS}^0 \ge 1$).
(b) Negative of the imaginary part of the intersubband
polarizability $- {\rm Im}\, \Pi_{\rm inter}$ in arbitrary units
as a function of frequency $\omega$ in the antiferromagnetic phase
(point $B_2$ in the phase diagram of Fig.~\protect\ref{fig:phase}(a)).
This quantity is related to the intensity of the intersubband
response in inelastic light scattering experiments.
As emphasized by the inset, the acoustic (Nambu-Goldstone) mode
dominates the low-$\omega$ intensity profile.
The spectrum is computed at $q / q_{\Delta} = 0.4$ and incorporates
a finite scattering rate $\gamma = 0.01 \Delta_{\rm SAS}^0$ to
simulate the effect of impurities;
the notation is the same as in Fig.~\protect\ref{fig:phase}.}
\label{fig:systematics}
\end{figure}

In the antiferromagnetic phase, the pattern of the
collective modes changes, as plotted in Fig.~\ref{fig:qw}.
The most obvious difference is the presence of an additional
region of particle-hole excitations at low frequencies.
This continuum arises because the wave functions in the antiferromagnetic
phase are linear combinations of wave functions from different
subbands,\cite{RS} which results in a mixing of intra- and
intersubband excitations.
A second feature in these spectra is the reappearance of an
optical spin-density excitation whose $q = 0$ frequency increases as
one goes deeper into the antiferromagnetic phase ($B_1 \rightarrow B_2$).
The dependence of the frequency of this mode, along with that of
the charge-density mode and the renormalized subband splitting,
is displayed in Fig.~\ref{fig:systematics}(a).
Finally, one observes a strong, linearly dispersing feature within
the lower particle-hole continuum.
This mode is the Nambu-Goldstone mode\cite{NG} arising from the
broken spin-rotation invariance in the antiferromagnetic phase.
A unique feature of this mode is that it is damped by
particle-hole excitations.

This damping leads naturally to the question of how much
spectral weight is associated with the Nambu-Goldstone mode.
This question has important implications for ILS
experiments,\cite{Decca,Plaut}
because the imaginary part of the polarizability is related to
the light scattering cross section.\cite{SDS,Raman}
In particular, the spin-density excitations
are directly observable in depolarized ILS, in which the polarization
of the scattered light is rotated by 90$^{\circ}$ with respect to the
incident light.\cite{Decca,Plaut}
We display in Fig.~\ref{fig:systematics}(b), therefore, the
imaginary part of the intersubband polarizability at point
$B_2$ of Fig.~\ref{fig:phase}(a) at a representative wave vector.
We see that the Nambu-Goldstone mode, the optical mode, and the
intersubband particle-hole continuum are all present, but that
the Nambu-Goldstone mode contains most of the spectral weight.
Since most ILS experiments are done at small
wave vector and moderate frequencies on this scale, this mode
may be difficult to observe.
In fact, the empirical signature of the new phase may simply be
the apparent disappearance of all ILS intensity.

In summary, we have computed the collective modes in antiferromagnetically
correlated quantum wells for a simple model within a conserving
approximation.
We find that two collective excitations in the intersubband spin-density
channel are associated with the antiferromagnetic phase, one optical
and one acoustic.
The acoustic excitation is the Nambu-Goldstone mode produced by the broken
spin symmetry in the antiferromagnetic state.
This mode is damped by particle-hole excitations but nonetheless
retains the majority of the spectral weight, indicating that most
of the inelastic light scattering intensity will be at low frequencies
in the antiferromagnetic phase.
The qualitative features of the collective excitations in this
symmetry-broken phase follow from very general principles, such as
the conservation of particle number, momentum, and energy\cite{conserving}
and Goldstone's theorem for phase transitions involving a spontaneously
broken continuous symmetry.\cite{NG}
Thus, while the quantitative details of our calculated collective mode
dispersion are undoubtedly model-specific, the qualitative features
should remain valid independent of our rather crude treatment
of the interaction term.

{\it Acknowledgements.}
The authors would like to thank P.~Littlewood for stimulating
discussion during the course of this work.
This work was supported by the ONR and the ARO.


\begin{references}
\bibitem{note1}Since we only examine the self-energy equations
for solutions which conserve 2D wave vector, spin- and
charge-density wave solutions normal to the layering direction
are explicitly excluded from these calculations.
\end{references}
\end{document}